\documentclass[aps,prl,showpacs,twocolumn]{revtex4}

\usepackage{amsmath,amssymb,graphicx}

\begin{document}
\noindent{\bf Comment on ``Accelerating cosmological expansion from shear and bulk viscosity''}\\
In a recent Letter \cite{one} the cause of the acceleration of the present Universe has been identified with
the shear viscosity of an imperfect relativistic fluid even in the absence of any bulk viscous contribution. If
 $\zeta$ and $\eta$ denote, respectively, the bulk and the shear viscosity coefficients,  
 the conclusion of Ref. \cite{one} is that the deceleration parameter can be negative 
 when $\zeta \to 0$ and $\eta \neq 0$. The gist of this comment is that the shear viscosity, if anything, 
 can only lead to an accelerated expansion over sufficiently small scales well inside 
 the Hubble radius. 

When the fluctuations of the metric are perturbative, as in Ref. \cite{one}, the gauge of the spatial gradients is given, in practice, by the product of the wavenumber $k$ and of the conformal time coordinate $\tau \simeq 1/{\mathcal H}$. 
Since $k\tau$ roughly equals the ratio between the Hubble radius and the typical spatial gradient, Ref. \cite{one} assumes that the Newtonian potentials are ${\mathcal O}({\mathcal H}^2/k^2)$, 
the velocity field is ${\mathcal O}({\mathcal H}/k)$ while the density (and its contrast) are both ${\mathcal O}(1)$.
We remark that this approximation scheme is tenable provided the relevant Fourier modes are located sufficiently inside the Hubble radius (i.e. $k \tau \gg 1$), as suggested in the previous paragraph. From this hierarchy all the equations easily follow if we set to zero the shift vector
in the four-dimensional space-time metric \cite{two}. On a more technical ground defining $u^2 = \gamma_{ij} u^{i} u^{j}$, the condition $g_{\mu\nu} u^{\mu} u^{\nu} =1$ for the four-velocity of the fluid implies $u^{0} = \sqrt{1 + u^2}/N$ which calls for an expansion in powers of $u^2$, for instance, in the explicit expressions of the covariant conservation equations \cite{one}.

The simplest way to proceed is to drop the assumption that the gravitational and hydrodynamic fluctuations are perturbatively small and study the effects of the bulk and shear viscosity within the expansion in spatial gradients. This nonperturbative scheme for the dynamics of the geometry and of the sources is the gravitational analog of companion methods employed in quantum field theory, in electrodynamics and in plasma physics \cite{three}. Order by order in the spatial gradients, the validity 
of the momentum and of the Hamiltonian constraints can be explicitly enforced. 
In the synchronous frame \cite{two} the nonlinear generalization of the deceleration parameter is given by $q(t, \vec{x}) = -1 
+ \dot{K}/\mathrm{Tr}K^2$ where $K_{ij} = - \dot{\gamma}_{ij}/2$ is the extrinsic curvature and the overdot 
denotes a derivation with respect to the cosmic time coordinate $t$. In the homogeneous 
and isotropic limit (i.e. $\gamma_{ij} \to a^2(t) \delta_{ij}$) we get $q(t, \vec{x}) \to - \ddot{a} a/\dot{a}^2$ since, by definition, 
$K = \gamma^{ij} K_{ij}$ and $\mathrm{Tr}K^2 = K_{i}^{j} K_{j}^{i}\geq 0$. From the $(00)$ component 
of the Einstein equations (contracted in the form $R_{\mu}^{\nu} = 8 \pi G [T_{\mu}^{\nu} - T \delta_{\mu}^{\nu}/2]$)
the equation obeyed by the deceleration parameter is:
\begin{equation}
q(t,\vec{x}) \mathrm{Tr}K^2 = 8 \pi G [ (\rho + 3 p)/2 + 3 \zeta K/2 + {\mathcal A}(t, \vec{x})],
\nonumber
\end{equation}
where $\rho$ and $p$ are, respectively, the energy density and the pressure of the fluid. To
lowest order in $u^2$, ${\mathcal A}(t,\vec{x})$ is given by $[ (\rho + p) u^2 + K \zeta u^2  + 2 \eta u^{k} u^{\ell} \overline{K}_{k\ell}]$ where
$  \overline{K}_{k\ell} = K_{k\ell}- K\gamma_{k\ell}/3$ denotes the traceless part of the extrinsic curvature \cite{four}.  
Thanks to the momentum constraint \cite{five} the peculiar velocity is proportional to a combination of gradients of the extrinsic curvature and of its trace. We conclude that to lowest order in the spatial gradients the sign of the deceleration 
parameter is solely determined by the bulk viscosity while the shear viscosity only appears to higher order
and it affects physical scales that are much smaller than $K^{-1}$ where $K^{-1}$ is, up to an irrelevant numerical factor, the inhomogeneous generalization of the Hubble radius. 
The nonperturbative results confirm that the sign of $q(t,\vec{x})$ is determined, to leading order, by the bulk viscosity coefficient \cite{six}. Over smaller scales the velocity field can be averaged in terms of a consistent coarsening  that respects both the Hamiltonian and the momentum constraints.
In summary, if the strong energy condition is enforced (e.g. $p=0$ and $\rho\geq 0$) whenever the bulk 
viscosity coefficient vanishes (i.e. $\zeta =0$) the deceleration parameter is positive semidefinite over the whole Hubble patch and  to leading order in the gradient expansion.  Thus, from the range of validity of the approximation scheme employed in \cite{one} and from the fully nonlinear analysis based on the gradient expansion, we conclude that the shear viscosity can only affect typical scales smaller than the Hubble radius.
The author wishes to thank S. Floerchinger and U. Wiedemann for useful exchanges of ideas.

\vspace{2mm}
\noindent Massimo Giovannini\\
{\it Department of Physics, 
Theory Division, CERN, 1211 Geneva 23, Switzerland  and INFN, Section of Milan-Bicocca, 20126 Milan, Italy}


\begin{thebibliography}{99}
\bibitem{two}  For consistency with the geometry adopted in \cite{one}
and for immediate convenience we shall require vanishing shift vector (i.e. $g_{0i} =0$) 
while $g_{00} = N^2$ and $g_{ij} = - \gamma_{ij}$. In the synchronous frame 
(i.e. $N=1$), $\gamma_{ij}$ carries exactly the $6$ degrees of freedom 
relevant for the subsequent discussion.
\bibitem{four} In the analysis the authors assume $\overline{K}_{ij}=0$ as it is clear from their expression of the deceleration parameter.
\bibitem{five} The explicit form of the momentum constraint is $\nabla_{i} K - \nabla_{k} K^{k}_{i} = 8 \pi G[ (\rho + p +\zeta K) u_{i} + 2 \eta \overline{K}_{i}^{k} u_{k} ]$ where $\nabla_{i}$ are the covariant derivatives defined in terms of $\gamma_{ij}$.
\end{thebibliography}
\end{document}